\documentclass[a4paper,11pt]{article}
\usepackage{tikz-feynman}
\usepackage{tikz}
\usepackage{float}
\usepackage{caption}
\usepackage{subcaption}
\usepackage{amsmath}
\usepackage{upgreek}
\usepackage{mathrsfs}
\usepackage{url}

\usepackage[pdftex]{hyperref}
\hypersetup{
  bookmarksopen=true, breaklinks=true, debug=true, %
  colorlinks=true, linkcolor=blue, citecolor=blue, urlcolor=blue
}

\title{Measurement of azimuthal modulations in SIDIS off proton target at COMPASS}

\author{Vendula Bene\v{s}ov\'{a}$^{a,*}$ on behalf of the COMPASS Collaboration}

\def\affiliation[a]{$^{a}$Charles University, Prague, Czech Republic}

\def\email{$^{*}$\href{mailto:vendula.benesova@cern.ch}{vendula.benesova@cern.ch}}

\newcommand\nukl{\mathrm{N}}
\newcommand\had{\mathrm{h}}
\newcommand\lep{\ell}
\newcommand\fin{\mathrm{X}}
\newcommand\kaon{\mathrm{K}}
\newcommand\qu{q}
\newcommand\PhT{P_\mathrm{T}}
\newcommand\pT{P_\perp}
\newcommand\khT{k_\mathrm{T}}

\newcommand\phih{\phi_h}
\newcommand\dd{\text{d}}
\newcommand\asin{A^{\sin \phih}_\text{LU}}
\newcommand\acos{A^{\cos \phih}_\text{UU}}
\newcommand\accos{A^{\cos 2\phih}_\text{UU}}

\newcommand\Gevc{GeV/$c$\ }

\def\abstract{The azimuthal angle ($\phih$) distribution of hadrons produced in deep inelastic scattering serves as a powerful tool for probing the nucleon structure in terms of transverse momentum dependent parton distribution functions and fragmentation functions. For an unpolarized nucleon, three azimuthal modulations arise: 
 $\cos\phih$ related to the Cahn effect, 
 $\cos2\phih$ linked to the Boer--Mulders function, and 
 $\sin\phih$ known as beam-spin asymmetry, 
each revealing insights into combinations of twist-two or higher-twist distribution and fragmentation functions.

The COMPASS collaboration at CERN collected semi-inclusive deep inelastic scattering events in 2016 and 2017 using a longitudinally polarized 160 \Gevc muon beam scattering off a liquid hydrogen target. Data from 2016 corresponding to about 1/3 of the full sample have been analyzed to measure the azimuthal modulations of charged hadrons. For the first time, the results were corrected for QED radiative effects using the DJANGOH MC generator.}

\def\FullConference{Presented on 11th April 2024 at 31st International Workshop on Deep Inelastic Scattering (DIS2024) in Grenoble, France.}

\makeatletter
    \def\@maketitle{%
  \newpage
  \null
  \vskip 2em%
  \begin{center}%
  \let \footnote \thanks
    {\LARGE \@title \par}%
    \vskip 1.5em%
    {\large
      \lineskip .5em%
      \begin{tabular}[t]{c}%
        \@author
      \end{tabular}\par}%
      \vskip .5em%
    {\large
      \lineskip .5em%
      \begin{tabular}[t]{c}%
        \affiliation[a]
      \end{tabular}\par}%
      \vskip .5em%
    {\large
      \lineskip .5em%
      \begin{tabular}[t]{c}%
        \email
      \end{tabular}\par}%
      \vskip 1em%
    {\small
      \lineskip .5em%
      \begin{tabular}[t]{c}%
      \parbox{0.8\textwidth}{\centering
        \FullConference}
      \end{tabular}\par}%
      \vskip 1em%
      \subsection*{Abstract}
    {\large
      \lineskip .5em%
      \parbox{0.8\textwidth}{
       \begin{flushleft}
      \begin{abstract}%
      \end{abstract}
      \end{flushleft}\par}}%
  \end{center}%
  \par
  \vskip 2em}
\makeatother

\begin{document}
\maketitle

\section{Theoretical introduction}

Semi-inclusive measurement of unpolarized DIS $\lep\nukl\rightarrow\lep'\had\fin$ is a powerful tool for studies of nucleon properties. Namely, the 3D nucleon structure is reflected in distributions of the azimuthal angle $\phih$ of final state hadrons $\had$, which can be described by the following cross-section formula~\cite{Bacchetta:2006tn}:
\begin{equation}\label{eq:cross-section}
     \frac{\dd^5\sigma}{\dd x\dd y \dd z \dd \phih \dd \PhT} = \sigma_0\biggl(1+ \varepsilon_1 \acos \cos \phih + \varepsilon_2 \accos \cos 2 \phih + \lambda \varepsilon_3  \asin \sin \phih \biggr)\ .
\end{equation}
\begin{figure}[tbp]
    \centering
    \begin{subfigure}[tbp]{0.24\textwidth}
    \scalebox{0.75}{
 \vspace{-25pt}\hspace{-45pt}\begin{tikzpicture}
  \begin{feynman}

    \coordinate (AA) at (2,-1.2);
    \coordinate (BB) at (4,-2.4);
    \coordinate (CC) at ($(BB) + (1,0.6)$);
    \coordinate (X) at ($(AA) + (2,1.2)$);
    \coordinate (DD) at ($(AA) + (1,0.6)$);
    \vertex (c1) at ($(DD)$);
    \vertex (c2) at ($(DD) - (1.25,-0.75)$);
    \coordinate (c3) at ($(DD) - (1.75,-1.05)+ (0.75,0.45)$);
    \coordinate (c4) at ($(DD) - (1.75,-1.05)- (0.75,0.45)$);
    \coordinate (EE) at ($(AA) + (0.4,-0.6) + (2,1.2)$);
    \coordinate (FF) at ($(BB) + (0.4,-0.6) + (2,1.2)$);
    \coordinate (EEE) at ($(AA) + (0.3,-0.45) + (1.75,1.05)$);
    \coordinate (FFF) at ($(BB) + (0.3,-0.45) + (1.75,1.05)$);
    \coordinate (GG) at ($(BB) + (1,0.6)$);
    \coordinate (HH) at ($(AA) + (1,0.6)$);
    \coordinate (A) at (0,0);
    \coordinate (B) at (4,-2.4);
    \coordinate (C) at ($(B) + (2,1.2)$);
    \vertex (az1) at ($(B) + (1.5,0.9)$);
    \coordinate (D) at ($(A) + (2,1.2)$);
    \fill[lightgray] (EE) -- (FF) -- (GG) -- (HH) -- cycle;    
    \fill[black] (EEE) -- (FFF) -- (GG) -- (HH) -- cycle;
    \draw[fill=white] (A) -- (B) -- (C) -- (D) -- cycle;
    \fill[black] (AA) -- (BB) -- (CC) -- (DD) -- cycle;

    \coordinate (E) at ($(AA) + (-0.4,0.6)$);
    \coordinate (F) at ($(BB) + (-0.4,0.6)$);
    \coordinate (G) at ($(BB) + (1,0.6)$);
    \coordinate (H) at ($(AA) + (1,0.6)$);
    \vertex (az2) at ($(BB) + (0.3,0.6)$);
    
    \fill[lightgray] (E) -- (F) -- (G) -- (H) -- cycle;
  
    \draw[->, line width=0.5pt] (0,-2) -- (1,-1.4) node[below]{$x$};
    \draw[->, line width=0.5pt] (0,-2) -- (0,-1) node[left]{$y$};
    \draw[->, line width=0.5pt] (0,-2) -- (1,-2.6) node[above]{$z$};
    \draw[->, line width=0.5pt, color=magenta] (c2) -- (c3) node[below right,font=\large]{$\boldsymbol{l'}$};
    \draw[->, line width=0.5pt, color=magenta] (c4) -- node[below,font=\large]{$\boldsymbol{l}$} (c2);
    \draw[boson, line width=0.5pt, color=green] (c2) -- (c1) node[pos=0.8, above,font=\large]{$\boldsymbol{q}$};
    \draw[->, line width=0.5pt, color=blue] (c1) -- node[pos=0.6, right,font=\large]{$\boldsymbol{P_h}$} (F); 
    \draw[->, line width=0.5pt, color=blue] (c1) -- node[pos=0.2, below,font=\large]{$\boldsymbol{\PhT}$} (E);
    \draw [->, bend right=45, color=blue]  (az1) to node[above,font=\large]{$\phih$} (az2);
    \draw [dashed, line width=.15pt] (DD) -- (X);
  \end{feynman}
\end{tikzpicture}}
\caption{~}\label{fig:GNS}
\end{subfigure}
    \begin{subfigure}[tbp]{0.21\textwidth}
    \scalebox{0.75}{
    \begin{tikzpicture}
  \begin{feynman}
    \vertex (i1) {\( \lep(l) \)};
    \vertex [below right=1.8cm of i1] (i2);
    \vertex [above right=1.3cm of i2] (i3) {\( \lep'(l') \)};
    \vertex [below =1.5cm of i2] (o1);
    \vertex [below left=of o1] (b) {\( \nukl(P) \)};
    \vertex [below right=2cm of o1] (a) {\( \)};
    \vertex [above=0.5cm of a] (a1) {\( \)};
    \vertex [above=0.5cm of a1] (a2) {\( \)};
    \vertex [above=1.5cm of a2] (a3) {\( \had(P_\had) \)};
    
    \diagram* {
      (i1) -- [fermion] (i2) -- [fermion] (i3),
      (i2) -- [boson, edge label = {\( \gamma^* \)}] (o1),
      (b) -- [fermion] (o1),
      (o1) -- [fermion] (a),
      (o1) -- [fermion] (a1),
      (o1) -- [fermion] (a2),
      (o1) -- [fermion] (a3),
    };
    \draw[fill=gray] (o1) circle [radius=0.25cm];
    \draw [decoration={brace}, decorate] (a2.north east) -- (a.south east)
          node [pos=0.5, right] {\(\fin\)};
  \end{feynman}
\end{tikzpicture}}
\caption{~}\label{fig:SIDIS}
\end{subfigure}
    \begin{subfigure}[tbp]{0.21\textwidth}
    \scalebox{0.75}{
    \begin{tikzpicture}
  \begin{feynman}
    \vertex (i1) {\( \lep(l) \)};
    \vertex [below=1cm of i1] (rad);
    \vertex [below right=1.8cm of i1] (i2);
    \vertex [above left=0.8cm of i2] (rad0);
    \vertex [above right=1.3cm of i2] (i3) {\( \lep'(l') \)};
    \vertex [below left=1.1cm of i3] (rad11);
    \vertex [left=0.1cm of rad11] (rad1);
    \vertex [below =1cm of i3] (rad2);
    \vertex [below =1.5cm of i2] (o1);
    \vertex [below left=of o1] (b) {\( \nukl(P) \)};
    \vertex [below right=2cm of o1] (a) {\( \)};
    \vertex [above=0.5cm of a] (a1) {\( \)};
    \vertex [above=0.5cm of a1] (a2) {\( \)};
    \vertex [above=1.5cm of a2] (a3) {\( \had(P_\had) \)};
    
    \diagram* {
      (i1) -- [fermion] (i2) -- [fermion] (i3),
      (i2) -- [boson, edge label = {\( \gamma^* \)}] (o1),
      (rad0) -- [boson, color=blue] (rad),
      (rad1) -- [boson, color=red] (rad2),
      (b) -- [fermion] (o1),
      (o1) -- [fermion] (a),
      (o1) -- [fermion] (a1),
      (o1) -- [fermion] (a2),
      (o1) -- [fermion] (a3),
    };
    \draw[fill=gray] (o1) circle [radius=0.25cm];
    \draw [decoration={brace}, decorate] (a2.north east) -- (a.south east)
          node [pos=0.5, right] {\(\fin\)};
  \end{feynman}
\end{tikzpicture}}
\caption{~}\label{fig:SIDIS2}
\end{subfigure}
    \begin{subfigure}[tbp]{0.21\textwidth}
    \scalebox{0.75}{
\begin{tikzpicture}
  \begin{feynman}
    \vertex (i1) {\( \lep(l) \)};
    \vertex [below=1cm of i1] (rad);
    \vertex [below right=1.8cm of i1] (i2);
    \vertex [above left=0.8cm of i2] (rad0);
    \vertex [above right=1.3cm of i2] (i3) {\( \lep'(l') \)};
    \vertex [below left=1.1cm of i3] (rad11);
    \vertex [left=0.1cm of rad11] (rad1);
    \vertex [below =1cm of i3] (rad2);
    \vertex [below =0.3cm of i2] (o11);
    \vertex [below =0.9cm of i2] (o111);
    \vertex [below =1.5cm of i2] (o1);
    \vertex [below left=of o1] (b) {\( \nukl(P) \)};
    \vertex [below right=2cm of o1] (a) {\( \)};
    \vertex [above=0.5cm of a] (a1) {\( \)};
    \vertex [above=0.5cm of a1] (a2) {\( \)};
    \vertex [above=1.5cm of a2] (a3) {\( \had(P_\had) \)};
    
    \diagram* {
      (i1) -- [fermion] (i2) -- [fermion] (i3),
      (i2) -- [boson] (o11),
      (o1) -- [boson] (o111),
      (rad0) -- [boson,quarter left, color=green] (rad1),
      (o11) -- [fermion,half right, color=magenta] (o111),
      (o111) -- [fermion,half right, color=magenta] (o11),
      (b) -- [fermion] (o1),
      (o1) -- [fermion] (a),
      (o1) -- [fermion] (a1),
      (o1) -- [fermion] (a2),
      (o1) -- [fermion] (a3),
    };
    \draw[fill=gray] (o1) circle [radius=0.25cm];
    \draw [decoration={brace}, decorate] (a2.north east) -- (a.south east)
          node [pos=0.5, right] {\(\fin\)};
  \end{feynman}
\end{tikzpicture}}
\caption{~}\label{fig:SIDIS3}
\end{subfigure}

    \caption{\textbf{(a)} Definition of the transverse momentum $\PhT$ and azimuthal angle $\phih$ of the final state hadron in the GNS. \textbf{(b)} Feynman diagram of the SIDIS process at the tree level. \textbf{(c,d)} Examples of $\mathscr{O}(\alpha^2)$ diagrams -- an illustration of QED radiative effects: initial-state radiation (blue), final-state radiation (red), vertex correction (green) and virtual photon self-energy (magenta). }
    \label{fig:GNSSIDIS}
\end{figure}
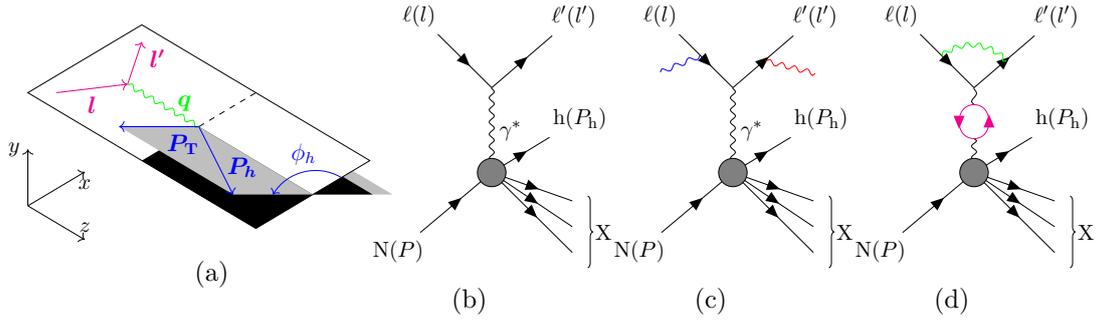
Apart from the standard (SI)DIS variables\footnote{inelasticity $y$, Bjorken variable $x$, fractional energy $z$ and transverse momentum $\PhT$ of the final state hadron} and azimuthal angle $\phih$, which is together with other hadronic variables defined in the $\upgamma^*$--N center-of-mass system (GNS) as shown in figure~\ref{fig:GNS}, the cross-section contains kinematical factors $\varepsilon_i$ and beam polarization $\lambda$. The amplitudes $A_\text{XU}^{f( \phih)}$ -- \textit{azimuthal asymmetries} -- are ratios of the structure functions $F_\text{XU}^{f( \phih)}$:
\begin{equation}
     A_\text{XU}^{f( \phih)}\bigl(x,z,\PhT^2,Q^2\bigr)\equiv\frac{F_\text{XU}^{f( \phih)}}{F_\text{UU}}\ , \qquad F_\text{UU}= F_\text{UU,T}+\varepsilon F_\text{UU,L}\ .
\end{equation}

Here, the subscript $\text{XU,Z}$ denotes polarization of the beam, the target and optionally the virtual photon ($\text{L}$ longitudinal, $\text{T}$ transverse and $\text{U}$ unpolarized). Assuming TMD factorization, the behaviour of the structure functions can be interpreted with the use of weighted (denoting the weight as $w(\boldsymbol{\khT},\boldsymbol{\pT})$) convolutions of TMD-PDFs $f^\qu(x,\khT^2,Q^2)$ and TMD-FFs\footnote{TMD-PDFs and TMD-FFs are abbreviations for transverse momentum dependent parton distribution functions and fragmentation functions. } $D^{\qu\rightarrow h} (z,\pT^2 ,Q^2)$~\cite{Bacchetta:2006tn}:
\begin{equation}
   F_\text{XU}^{f( \phih)}=\mathscr{C}[w fD]= x \sum_\qu e^2_\qu\int\dd^2 \boldsymbol{\khT}\dd^2\boldsymbol{\pT}\delta^{(2)}(z\boldsymbol{\khT}+\boldsymbol{\pT}-\boldsymbol{\PhT})w f^\qu D^{\qu\rightarrow h}\ .
\end{equation}
If we restrict ourselves to the leading twist TMDs, neglect quark-gluon-quark correlations and in the case of $F_\text{UU}^{\cos \phih}$ utilize the Wandzura--Wilczek-type approximation~\cite{Bastami_2019}, only two non-zero structure functions remain for unpolarized SIDS (denoting $\ \boldsymbol{\hat{h}}=\boldsymbol{\PhT}/|\boldsymbol{\PhT}|$):
\begin{equation}
\begin{aligned}
     F^{\cos 2 \phih}_\text{UU} &= \mathscr{C}\left[\frac{2(\boldsymbol{\hat{h}}\cdot \boldsymbol{\khT})(\boldsymbol{\hat{h}}\cdot \boldsymbol{\pT})-(\boldsymbol{\khT}\cdot\boldsymbol{\pT})}{zMM_h}h_1^\perp H_1^\perp\right]\ ,\\
    F^{\cos  \phih}_\text{UU} &= \frac{2M}{Q}\mathscr{C}\left[-\frac{(\boldsymbol{\hat{h}}\cdot \boldsymbol{\khT})}{M}f_1D_1+\frac{\khT^2(\boldsymbol{\hat{h}}\cdot \boldsymbol{\pT})}{zM^2M_h}h_1^\perp H_1^\perp\right],
\end{aligned}
\label{eq:TMDs}
\end{equation}
where $f_1$ is the unpolarised and $h_1^\perp$ the Boer--Mulders TMD-PDF, while $D_1$ is the unpolarised and $H_1^\perp$ the Collins TMD-FF.

\section{Data analysis}

The preliminary results of the 2016 liquid hydrogen data at COMPASS have been presented at previous conferences explaining already many steps of the analysis~\cite{Moretti:2020proc, Matousek:2019dlk}. Newly, the data sample (both measured and simulated) has been extended and QED radiative effects have been accounted for up to the order of $\alpha^2$.

\subsection{Treatment of background from diffractive vector meson decays}\label{sec:dvms}

\begin{figure}[tbp]
    \centering
    \includegraphics[width=.35\textwidth]{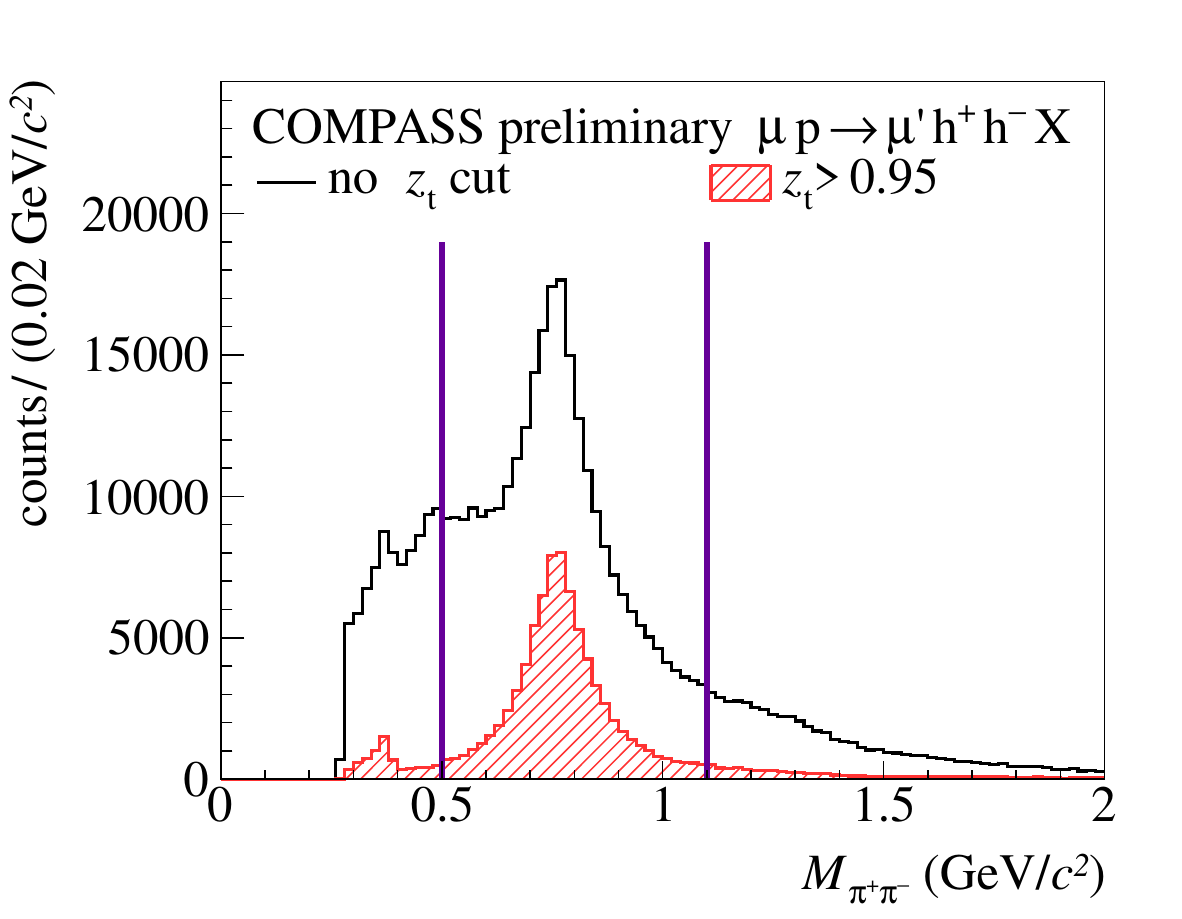}
    \includegraphics[width=.35\textwidth]{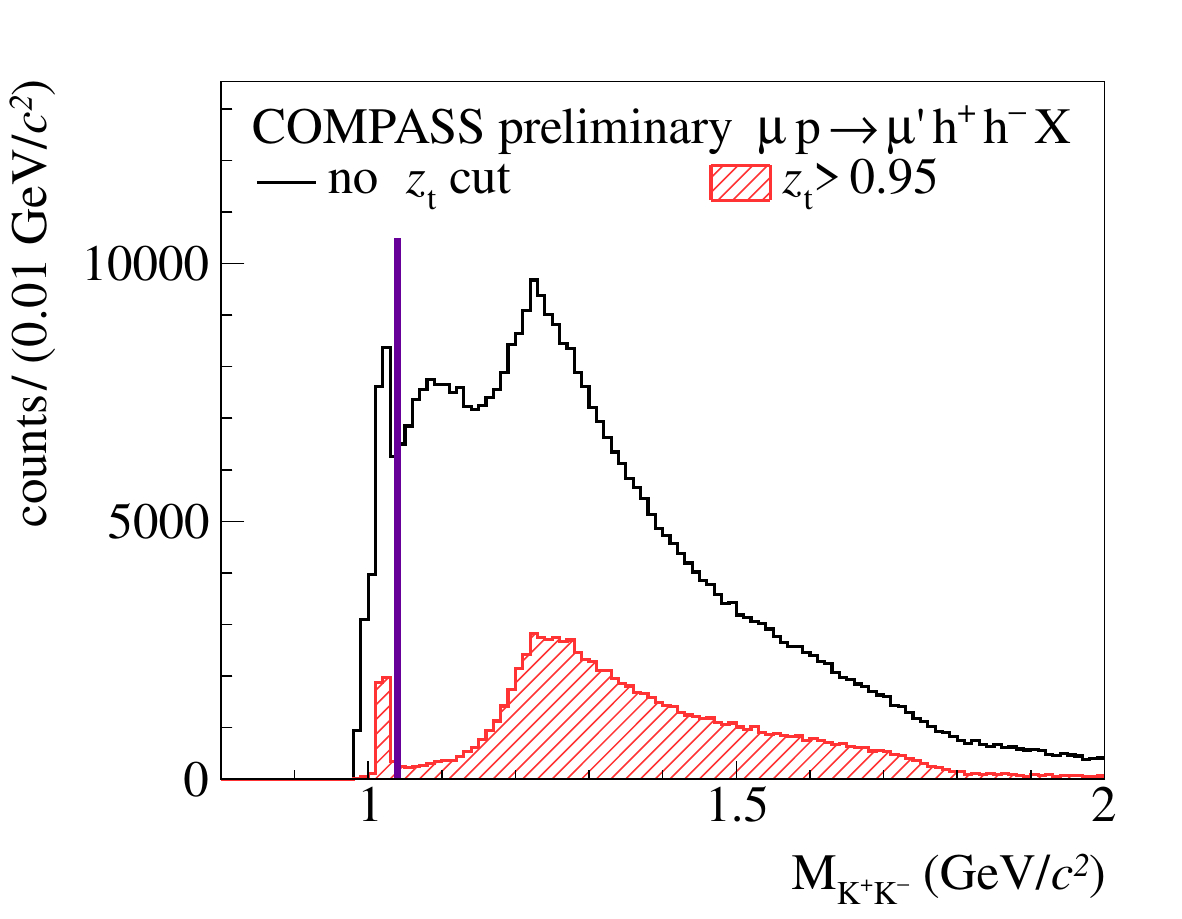}
    \caption{The invariant mass distributions for (left) $\uprho^0\rightarrow\uppi^+\uppi^-$ and (right) $\upphi\rightarrow\kaon^+\kaon^-$ decay hypotheses. }
    \label{fig:invm}
\end{figure}

\begin{figure}[tbp]
    \centering
    \includegraphics[width=.37\textwidth]{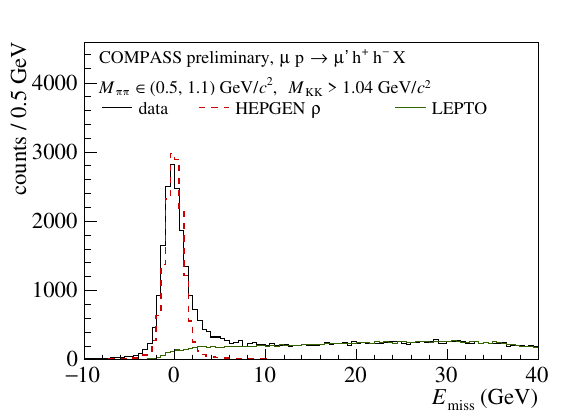}
    \includegraphics[width=.37\textwidth]{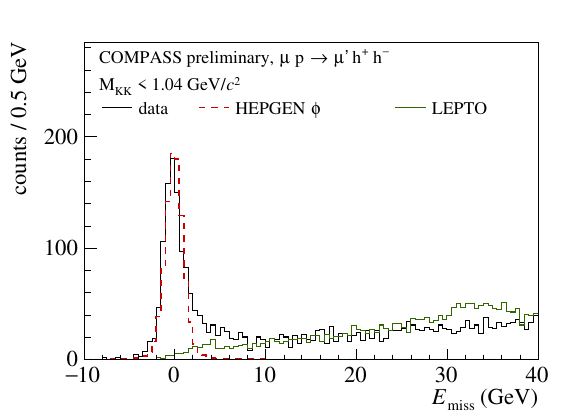}
    \caption{The missing mass distributions for (left) $\uprho^0\rightarrow\uppi^+\uppi^-$ and (right) $\upphi\rightarrow\kaon^+\kaon^-$ candidates. }
    \label{fig:emiss}
\end{figure}

Hadrons produced in hard exclusive processes are a background to the SIDIS measurements. The only significant contributors are the decays $\uprho^0(770)\rightarrow\uppi^+\uppi^-$ and $\upphi(1020)\rightarrow\kaon^+\kaon^-$~\cite{Agarwala_2020}, which are visible in the invariant mass distributions in figure~\ref{fig:invm}. Two procedures ensure the elimination of the background from diffractive vector mesons (DVM). 
First, in the case of reconstructing both charged hadrons in the spectrometer, one can apply the following constraint, which is a result of energy conservation, on $z_\text{t}\equiv z_{\had^+}+z_{\had^-}<0.95$ (referred to as \textit{DVM cut}).
The cases in which only one of the decay products is reconstructed are simulated by the HEPGEN MC generator~\cite{sandacz2012hepgengeneratorhard}.  Their azimuthal distributions are scaled according to the size of the peak in the missing energy histograms shown in figure~\ref{fig:emiss} and subtracted from the measured distributions of the data (\textit{DVM subtraction}).

\subsection{Radiative correction}
  The TMD framework outlined in the formulae~\ref{eq:cross-section}--\ref{eq:TMDs} does not account for QED radiative effects depicted in figure~\ref{fig:SIDIS2} and~\ref{fig:SIDIS3}. In particular, the real photon emission affects the measured leptonic DIS variables as well as the hadronic variables. An approach to this problem based on the DJANGOH~\cite{Charchula:1994kf} MC generator has been recently adopted by COMPASS~\cite{Stolarski2023}. We correct the number of events in each $\phih$ bin by the ratio of events simulated in that bin by DJANGOH with radiative effects on and off (normalized to the MC luminosity).
\section{Results and conclusion}

After applying the DVM and radiative corrections and after correcting for experimental acceptance, the azimuthal distributions are fitted with equation~\ref{eq:cross-section}, extracting the asymmetries. The one-dimensional dependences on $x$, $z$ and $\PhT$ are presented in the following figures; a multidimensional analysis is currently being finalized for publication.

\begin{figure}[tbp]
    \centering
    \includegraphics[width=.49\textwidth]{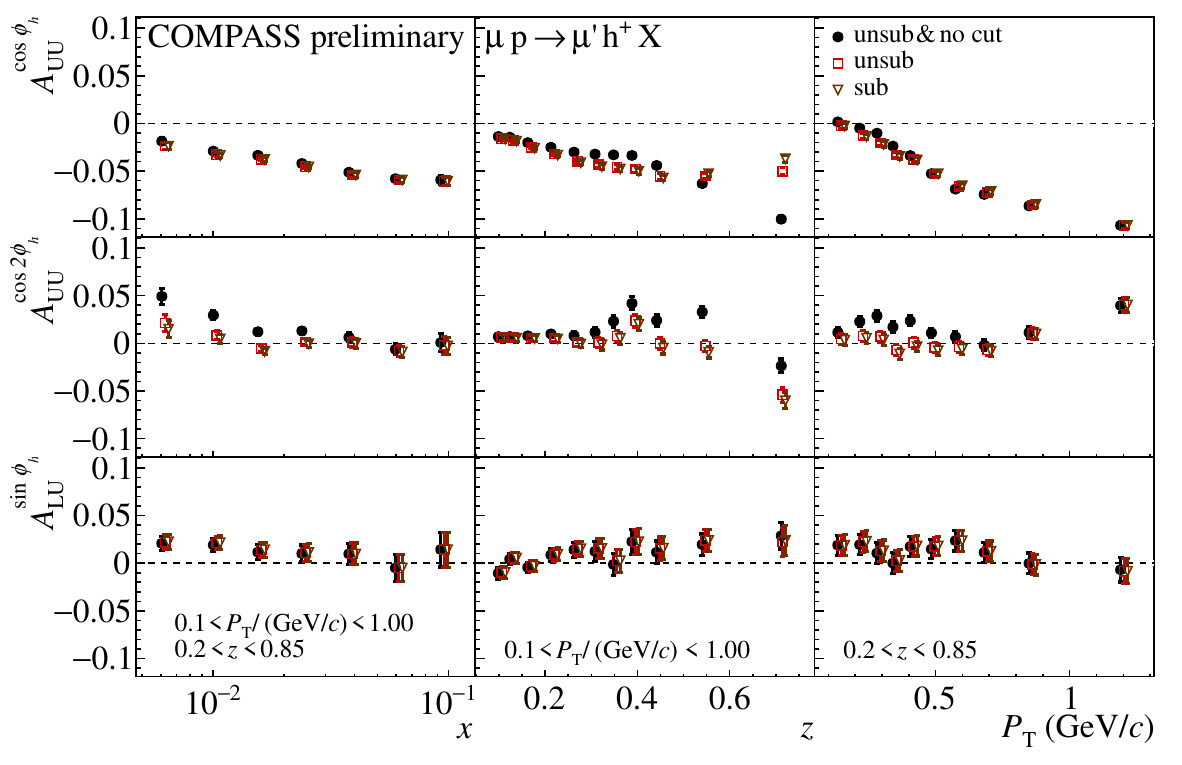}
    \includegraphics[width=.49\textwidth]{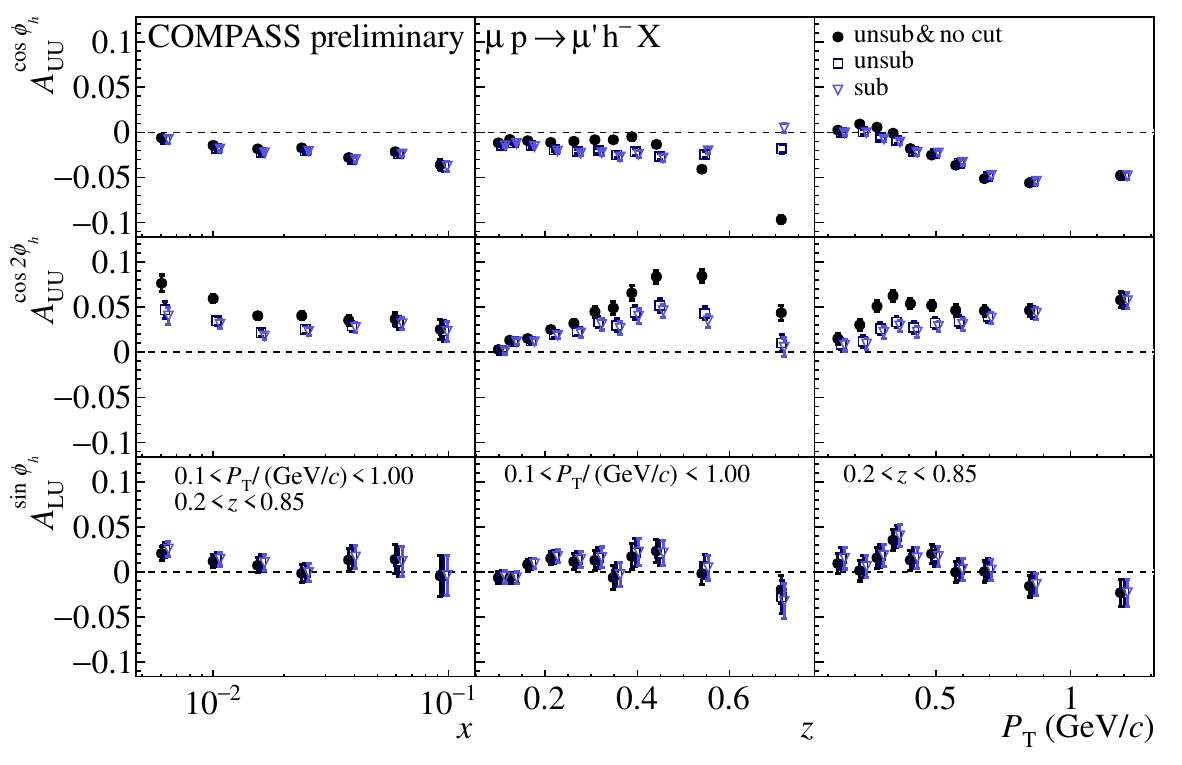}
    
    \caption{The comparison of the azimuthal asymmetries before applying DVM cut (unsub \& no cut), after applying the cut (unsub) and after DVM subtraction (sub) for (left) $\had^+$ and (right) $\had^-$.
    Note, that the majority of the background is removed by the DVM cut.
    The radiative correction is not applied.}
    \label{fig:compdvm}
\end{figure}

\begin{figure}[tbp]
    \centering
    \includegraphics[width=.49\textwidth]{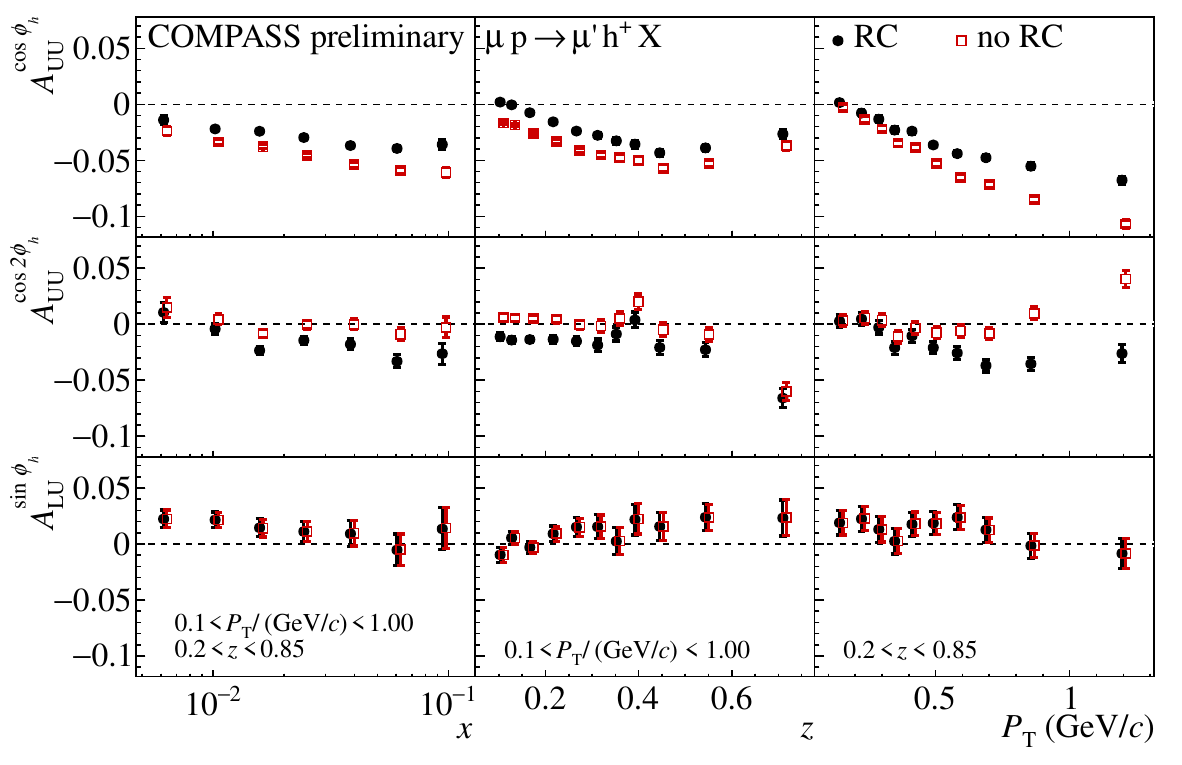}
    \includegraphics[width=.49\textwidth]{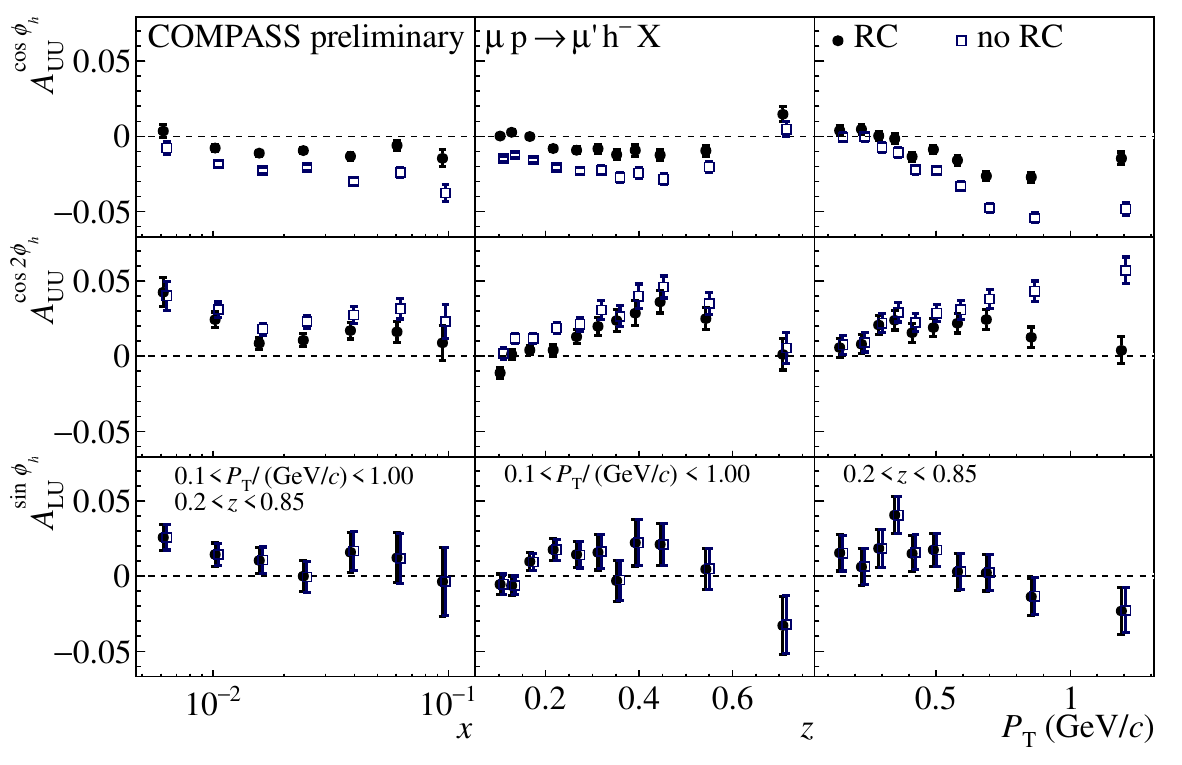}
    
    \caption{The comparison of the azimuthal asymmetries with and without applied radiative correction for (left) $\had^+$ and (right) $\had^-$.
    Note the qualitative change for $\accos$ for $\had^+$, which becomes clearly negative.}
    \label{fig:comprad}
\end{figure}

\begin{figure}[tbp]
    \centering
    \includegraphics[width=.58\textwidth]{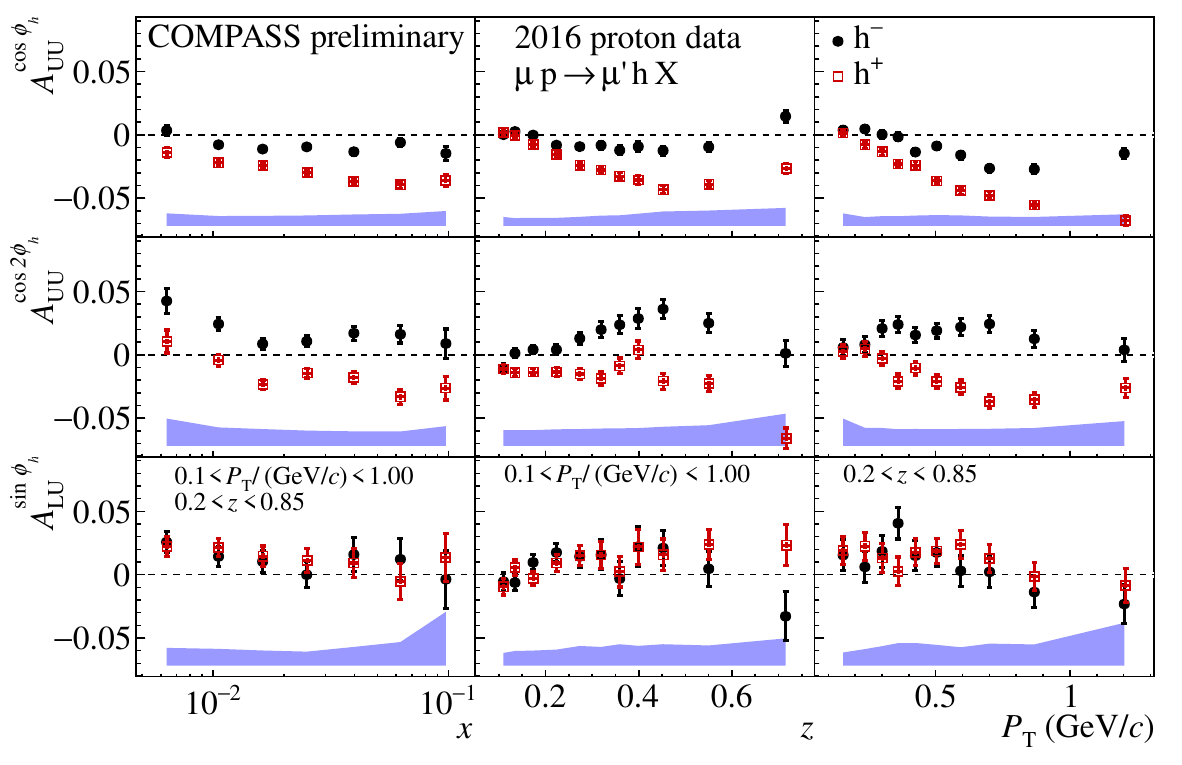}
    \caption{The final results for azimuthal asymmetries in their 1D $x$, $z$ and $\PhT$ dependence. The blue band represents systematic uncertainty and is common for $\had^+$ and $\had^-$. }
    \label{fig:res}
\end{figure}

Figure~\ref{fig:compdvm} demonstrates the effect of the background hadrons from DVMs as explained in section~\ref{sec:dvms}. The high $z$ bins, together with the low $x$ and $\PhT$ bins, are the most affected by the background. Figure~\ref{fig:comprad} shows the effect of the radiative corrections on the azimuthal asymmetries. The lower $z$ bins together with the higher $x$ and $\PhT$ bins are affected the most by this correction. No effect is observed for $\asin$.
The final results are shown in figure~\ref{fig:res}. The systematic uncertainty of the final results includes four contributions (from acceptance, DVM subtraction, radiative correction and period compatibility) evaluated for each bin separately. As expected from the expressions \ref{eq:TMDs}, asymmetries $\acos$ and $\accos$ show strong kinematical dependences, while $\asin$ is close to compatibility with zero. The negative trend of $\acos$ is related to the Cahn effect \cite{Cahn:1978se}.

\section*{Acknowledgements}
The work of the author has been supported by the 
Charles University grant PRIMUS/22/SCI/017.

\bibliographystyle{utphys}
\bibliography{bibliography}

\providecommand{\href}[2]{#2}\begingroup\raggedright\begin{thebibliography}{1}

\bibitem{Bacchetta:2006tn}
A.~Bacchetta {\em et~al.}, ``{Semi-inclusive deep inelastic scattering at small transverse momentum},'' \href{http://dx.doi.org/10.1088/1126-6708/2007/02/093}{{\em JHEP} {\bfseries 02} (2007) 093}, \href{http://arxiv.org/abs/hep-ph/0611265}{{\ttfamily arXiv:hep-ph/0611265}}.

\bibitem{Bastami_2019}
S.~Bastami {\em et~al.}, ``{Semi-Inclusive Deep Inelastic Scattering in Wandzura--Wilczek-type approximation},'' \href{http://dx.doi.org/10.1007/JHEP06(2019)007}{{\em JHEP} {\bfseries 06} (2019) 007}, \href{http://arxiv.org/abs/1807.10606}{{\ttfamily arXiv:1807.10606 [hep-ph]}}.

\bibitem{Moretti:2020proc}
A.~Moretti (COMPASS), ``{Azimuthal asymmetries of hadrons produced in unpolarized SIDIS at COMPASS},'' \href{http://dx.doi.org/10.1088/1742-6596/1435/1/012043}{{\em J. Phys. Conf. Ser.} {\bfseries 1435} no.~1, (2020) 012043}.

\bibitem{Matousek:2019dlk}
J.~Matou\v{s}ek (COMPASS), ``{Measurement of the azimuthal modulations of hadrons in unpolarised SIDIS},'' \href{http://dx.doi.org/10.22323/1.352.0189}{{\em PoS} {\bfseries DIS2019} (2019) 189}, \href{http://arxiv.org/abs/1907.08851}{{\ttfamily arXiv:1907.08851 [hep-ex]}}.

\bibitem{Agarwala_2020}
J.~Agarwala {\em et~al.} ({COMPASS}), ``Contribution of exclusive diffractive processes to the measured azimuthal asymmetries in {SIDIS},'' \href{http://dx.doi.org/10.1016/j.nuclphysb.2020.115039}{{\em Nuclear Physics B} {\bfseries 956} (July, 2020) 115039}.

\bibitem{sandacz2012hepgengeneratorhard}
A.~Sandacz and P.~Sznajder, ``{HEPGEN -- generator for hard exclusive leptoproduction}.'' 2012.
\newblock \href{http://arxiv.org/abs/1207.0333}{{\ttfamily arXiv:1207.0333 [hep-ph]}}.

\bibitem{Charchula:1994kf}
K.~Charchula {\em et~al.}, ``{Combined QED and QCD radiative effects in deep inelastic lepton - proton scattering: The Monte Carlo generator DJANGO6},'' \href{http://dx.doi.org/10.1016/0010-4655(94)90086-8}{{\em Comput. Phys. Commun.} {\bfseries 81} (1994) 381--402}.

\bibitem{Stolarski2023}
M.~Stolarski ({COMPASS}), ``{COMPASS} results on pion and kaon multiplicities from {SIDIS} on proton target.'' 2023.
\newblock \url{https://wwwcompass.cern.ch/compass/publications/conf_proc/proc/t2023/menu2023_stolarski.pdf}.
\newblock To appear in proceedings of {MENU} 2023.

\bibitem{Cahn:1978se}
R.~N. Cahn, ``{Azimuthal Dependence in Leptoproduction: A Simple Parton Model Calculation},'' \href{http://dx.doi.org/10.1016/0370-2693(78)90020-5}{{\em Phys.\ Lett.\ B} {\bfseries 78} (1978) 269--273}.

\end{thebibliography}\endgroup


\providecommand{\href}[2]{#2}\begingroup\raggedright\endgroup
\end{document}